\documentclass[twocolumn,astrosymb,tighten,twocolappendix]{aastex631}

\usepackage{graphicx}    % 画像を扱うためのパッケージ
\usepackage{afterpage}   % 図を次のページに配置するためのパッケージ

\usepackage{CJKutf8}%! To show the Japanese language.

\received{Jan 29, 2025}
\revised{Feb 11, 2025}
\accepted{Feb 22, 2025}

\graphicspath{{./}{./figures/}}

\newcommand{\Menc}{M_{\rm enc}}
\newcommand{\Mps}{M_{\rm ps}}
\newcommand{\Mpsest}{M_{\rm ps,est}}
\newcommand{\tps}{t_{\rm ps}}
\newcommand{\rrw}{r_{\rm rw}}
\newcommand{\rkepler}{r_{\rm Kepler}}
\newcommand{\vrot}{v_{\rm rot}}
\newcommand{\vkepler}{v_{\rm Kepler}}
\newcommand{\vkeplerps}{v_{\rm Kepler,ps}}
\newcommand{\vkeplertot}{v_{\rm Kepler,tot}}
\newcommand{\vinfall}{v_{\rm infall}}
\newcommand{\vff}{v_{\rm ff}}
\newcommand{\vffps}{v_{\rm ff,ps}}
\newcommand{\vfftot}{v_{\rm ff,tot}}
\newcommand{\fkepler}{f_{\rm K}}
\newcommand{\fff}{f_{\rm ff}}
\newcommand{\K}{\rm K}
\newcommand{\cs}{c_{\rm s}}

\newcommand{\msun}{M_\odot}

\newcommand{\cc}{{\rm cm^{-3}}}
\newcommand{\kms}{{\rm km\,s^{-1}}}

\shorttitle{Origins of the factor-of-three problem}
\shortauthors{Hirano et al.}

\begin{document}

\begin{CJK}{UTF8}{min}%! To show the Japanese language.

%%%%%%%%%%%%%%%%%%%%%%%%%%%%%%%%%%%%%%%%%%%%%%%%%%%%%%%%%%%%%%%%%%%%%%%%%%%%%%%%

\title{
Velocity Structure of Circumstellar Environment around Class 0/I Protostars: \\
Uncertainty in the Protostellar Mass Estimation Using Circumstellar Velocities}
\correspondingauthor{Shingo Hirano}
\email{shingo-hirano@kanagawa-u.ac.jp}

\author[0000-0002-4317-767X]{Shingo Hirano}
\affiliation{Department of Applied Physics, Faculty of Engineering, Kanagawa University, Kanagawa 221-0802, Japan}
\affiliation{Department of Astronomy, School of Science, University of Tokyo, Tokyo 113-0033, Japan}

\author[0000-0003-3283-6884]{Yuri Aikawa}
\affiliation{Department of Astronomy, School of Science, University of Tokyo, Tokyo 113-0033, Japan}

\author[0000-0002-0963-0872]{Masahiro N. Machida}
\affiliation{Department of Earth and Planetary Sciences, Faculty of Science, Kyushu University, Fukuoka 819-0395, Japan}

%%%%%%%%%%%%%%%%%%%%%%%%%%%%%%%%%%%%%%%%%%%%%%%%%%%%%%%%%%%%%%%%%%%%%%%%%%%%%%%%

%% Abstract: =< 250 words
\begin{abstract}
Recent high-resolution observations have enabled detailed investigations of the circumstellar environments around Class 0/I protostars.
Several studies have reported that the infall velocity of the envelope is a few times smaller than the free-fall velocity inferred from protostellar masses estimated via the observed rotational velocity of their Keplerian disks.
To explore the physical origins of the slow infall, we perform a set of three-dimensional resistive magnetohydrodynamic simulations of the star formation process, extending to $10^5\,$yr after protostar formation.
Our simulations show that the infall velocity decreases markedly at the outer edge of the pseudo-disk (at radii of $\sim\!100-1000\,$au) and is much slower than the expected free-fall velocity.
The degree of this reduction depends on (1) the initial magnetic field strength, (2) the alignment between the initial field and the rotation axis, and (3) the evolutionary stage of the system.
Across our parameter space, the ratio of the infall velocity to the free-fall velocity is as small as $0.2-0.5$, which is consistent with the observations.
We further examine the reliability of protostellar mass estimates derived from infall and rotational velocities.
While the mass derived from disk rotation closely matches the true value, deviation by a factor of $0.3-2$ is found for the estimates using the infall velocity; it is underestimated due to slow infall, but could also be overestimated due to the contribution of disk mass.
These findings underscore the critical role of magnetic fields in shaping star formation dynamics and highlight the uncertainties associated with protostellar mass estimates.
\end{abstract}

\keywords{Magnetohydrodynamical simulations (1966) --- Star formation (1569) --- Protostars (1302) --- Circumstellar disks (235) --- Protoplanetary disks (1300) --- Circumstellar envelopes (237)}

%%%%%%%%%%%%%%%%%%%%%%%%%%%%%%%%%%%%%%%%%%%%%%%%%%%%%%%%%%%%%%%%%%%%%%%%%%%%%%%%

%%%%%%%%%%%%%%%%%%%%%%%%%%%%%%%%%%%%%%%%%%%%%%%%%%%%%%%%%%%%%%%%%%%%%%%%%%%%%%%%
\section{Introduction} \label{sec:intro}
%%%%%%%%%%%%%%%%%%%%%%%%%%%%%%%%%%%%%%%%%%%%%%%%%%%%%%%%%%%%%%%%%%%%%%%%%%%%%%%%

Recent advancements in observational facilities such as ALMA and JWST have enabled detailed studies of the complex structures around protostars, including the protostar itself, Keplerian disks, infalling-rotating envelopes, outflows/jets, and streamers \citep[e.g.,][]{, Tokuda2014, Tokuda2018, ALMA2015, Tobin2020, Pineda2023}.
ALMA Large programs, for example, have provided invaluable insights into the star and planet formation process \citep[e.g.,][]{Andrews2018DSHARP-I, Ohashi2023eDisk-I}.
The use of multiple molecules now allows us to probe the velocity structures of the circumstellar environment and clarify how its various components interact \citep[see the schematic summary in Figure~11 of][]{Tychoniec2021}.

\begin{figure}[t!]
\includegraphics[width=1.0\linewidth]{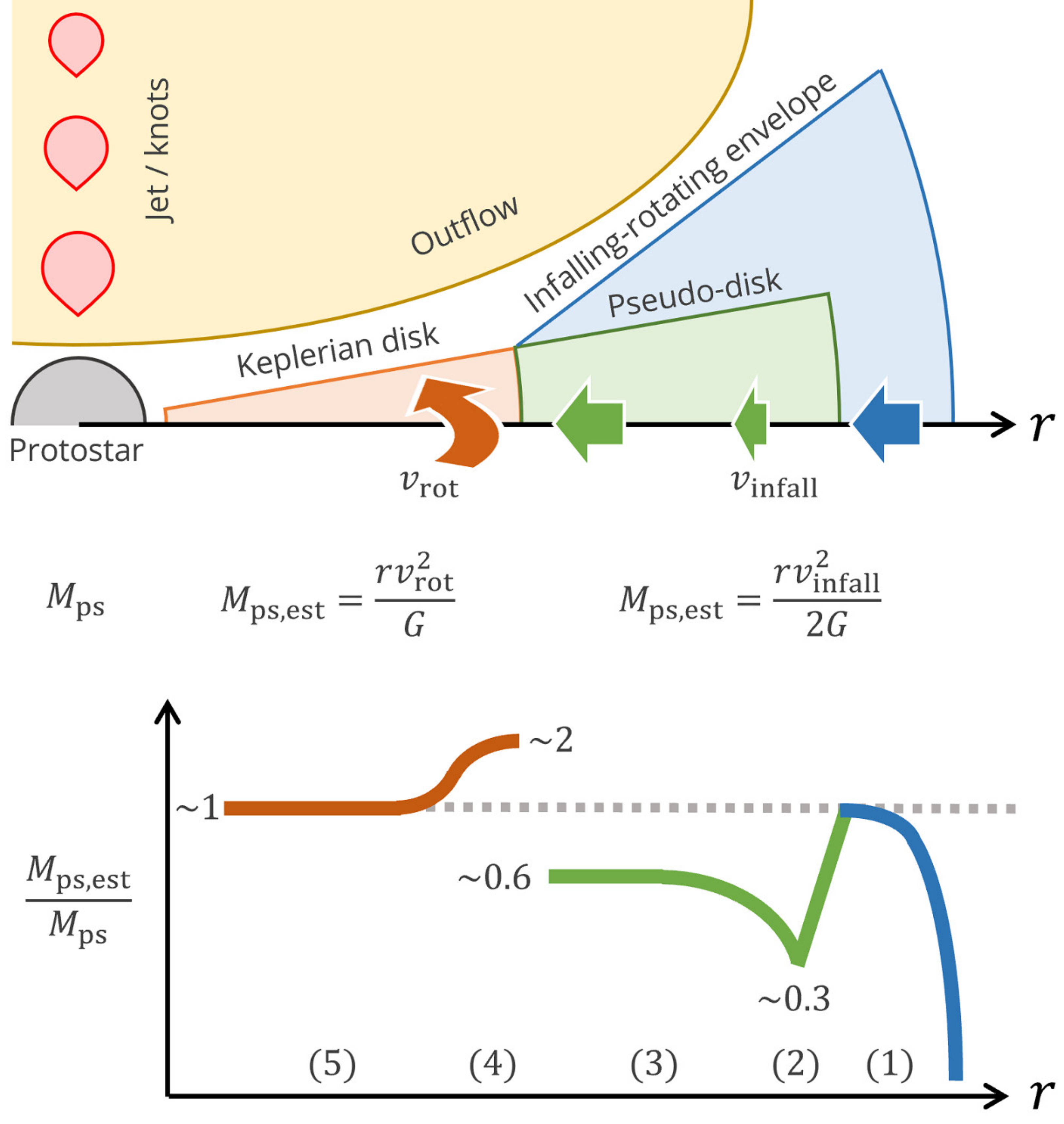}
\caption{
A schematic overview of the velocity structure around Class 0/I protostars, derived from the analysis of a set of simulations in this work.
(Top) The infall velocity of the infalling-rotating envelope is lower than the free-fall velocity, with significant deceleration occurring at the outer edge of the pseudo-disk (see also Figure~\ref{fig:2dmap_T00}).
The rotationally-supported disk rotates at approximately the Keplerian velocity.
(Bottom) The central protostellar mass estimated using infall (blue and green) and rotation (red) velocities ($\Mpsest$) differs from the true protostellar mass ($\Mps$) by a factor of up to several (see also Figure~\ref{fig:Mps_est-true}).
The circumstellar region is divided into 5 regions based on velocity structures.
}
\label{fig:overview}
\end{figure}

The mass and age of a protostar are the essential parameters to study the star formation process.
While they can be determined from the stellar spectrum using the Hertzsprung-Russell diagram, there are uncertainties due to the variations of protostellar evolution models \citep[e.g.][]{Simon2000}.
Therefore, it is important to estimate the protostar mass $\Mps$ using the velocity structure of the circumstellar gas.
A common approach assumes that the observed velocity at a given radius reflects either the disk's Keplerian rotation or the envelope's free-fall velocity.
If a Keplerian disk is spatially resolved, the protostellar mass is estimated as $\Mps = r\vrot^2/G$ assuming the rotation velocity equals the Keplerian velocity $\vrot = \vkepler = \sqrt{G\Mps/r}$, where $G$ is the gravitational constant and $r$ is the radius from the protostar.
On the other hand, the infall velocity is assumed to equal the free-fall velocity $\vinfall = \vff = \sqrt{2G\Mps/r}$, yielding $\Mps = r\vinfall^2/(2G)$ (see the top panel of Figure~\ref{fig:overview}).
Thus, the protostellar mass can be derived from the rotational velocity when a disk is spatially resolved, while the infall velocity is used, otherwise.

With the improved sensitivity and spatial resolution, it has become possible to directly observe Class 0/I protostars at very early evolutionary stages ($\Mps \lesssim 0.1\,\msun$) \citep[e.g., the eDisk program;][]{Ohashi2023eDisk-I}.
At such young stages, detecting a well-defined Keplerian disk can be challenging, making envelope properties such as $\vinfall$ crucial for mass estimation.
Analytic models, notably the Cassen \& Moosman formulation \citep{CassenMoosman1981}, have guided these efforts.

However, a persistent discrepancy has emerged; the protostellar mass estimated from infall velocities tends to be lower than that estimated from the Keplerian rotation velocity.
In other words, when the mass inferred from the Keplerian rotation velocity is used as a reference, the observed infall velocity is slower by a factor of $\sim 3$ compared to the free-fall velocity ($\vff$) predicted based on that mass assumption.
We call this a ``factor-of-three problem'' hereafter \citep{Ohashi2014}.
This has been reported in several sources: L1527~IRS \citep[$\vff/\vinfall \sim 3$;][]{Ohashi2014}, L1551~NE \citep[$\sim\!4$;][]{Takakuwa2013}, L1551~IRS~5 \citep[$\sim\!4$; ][]{Chou2014, Takakuwa2020}, TMC-1A \citep[$\sim\!3$;][]{Aso2015}, and L1489~IRS \citep[$\sim\!2. 5$;][]{Sai2022}.
In these cases, the mass estimates from infall velocities alone can be $0.1-0.25$ times the mass estimates based on rotation velocity.
This discrepancy raises important questions: {\it Is this factor-of-three problem universal? What is its origin, and what is the typical ratio of $\vinfall/\vff$ at different evolutionary stages?}
Possible causes include angular momentum conservation, magnetic braking, turbulence, non-axisymmetric structures, and observational limitations (e.g., spatial resolution and interpretation).

%%%%%%%%%%%%%%%%%%%%%%%%%%%%%%%%%%%%%%%%
%%%%%%%%%%%%%%%%%%%%%%%%%%%%%%%%%%%%%%%%

Over the past few decades, theoretical and numerical studies have played a pivotal role in advancing the understanding of star formation.
Magnetohydrodynamic (MHD) simulations, in particular, have become a standard tool for investigating how a gravitationally collapsing cloud evolves into a protostar and its circumstellar material \citep[e.g.,][]{MachidaBasu2019}.
A key advantage of such simulations is their ability to probe the earliest stages of star formation, just before and immediately after protostar formation, which remains difficult to observe directly due to high extinction and limited spatial resolution.
By self-consistently incorporating magnetic fields, non-ideal MHD effects, and increasingly sophisticated microphysics, these models have demonstrated that magnetic fields play a crucial role in regulating angular momentum transport, disk formation, and disk size \citep[e.g.,][]{Tsukamoto2023PP7}.
Understanding magnetic effects has emerged as a central problem in current star formation theory, also in the velocity structure \citep[e.g.,][]{Aso2015, Sai2022}.

\citet{Hirano2020} performed a set of resistive MHD simulations but focused on the early stages ($5000\,$yr after protostar formation), compared with the typical age of Class~0 ($\sim\!10^4\,$yr) and Class~I ($\sim\!10^5\,$yr) objects.
Longer-term simulations are needed to fill the gap between theoretical and observational studies.
Additionally, the influence of misalignment between the rotation axis and the magnetic field direction in prestellar clouds is becoming increasingly relevant, as recent observations and simulations suggest that it may impact various aspects of star formation \citep[e.g.,][]{Galametz2020, Huang2024, KinoshitaNakamura2023}.\footnote{However, some observational studies report weak or no correlation between misalignment and observed properties \citep{Yen2021, Gupta2022}.}

The present study uses resistive MHD simulations to follow the dynamical evolution of the circumstellar envelope and disk formation.
We specifically investigate velocity distributions of the circumstellar environment: $\vinfall$, $\vrot$, $\vinfall/\vff$, and $\vrot/\vkepler$.
At the outer edge of the pseudo-disk (radii $\sim100-1000\,$au), the infall velocity naturally decreases to about $0.3-0.4$ times the free-fall velocity.
This reduction explains why mass estimates based on $\vinfall$ can be much smaller than the true protostellar mass (see the bottom panel of Figure~\ref{fig:overview}). 
We also examine how the factor-of-three discrepancy depends on physical parameters, including the initial magnetic field strength, the angle between the field and rotation axes, and the evolutionary stage.
Our results provide a theoretical framework for interpreting future observations of Class 0/I objects, especially those without detectable Keplerian disks.

%%%%%%%%%%%%%%%%%%%%%%%%%%%%%%%%%%%%%%%%%%%%%%%%%%%%%%%%%%%%%%%%%%%%%%%%%%%%%%%%
\section{Methods} \label{sec:methods}
%%%%%%%%%%%%%%%%%%%%%%%%%%%%%%%%%%%%%%%%%%%%%%%%%%%%%%%%%%%%%%%%%%%%%%%%%%%%%%%%

We perform a set of three-dimensional resistive MHD simulations of the star formation process in which only ohmic dissipation is considered as the dissipation term of the magnetic field \citep{Machida2007}.
We start from a gravitationally unstable gas cloud \citep[as in][]{Hirano2020} to investigate the velocity distribution of the circumstellar gas of a Class 0/I protostar.
A nested-grid code is used to solve the resistive MHD equations under a barotropic equation of state.
We set the size of the computational domain $2\times10^5\,$au and the finest grid $0.38\,$au.
This computational domain is large enough to cover the entire circumstellar environment, and the finest grid size is small enough to numerically resolve the velocity structure of the circumstellar disk and envelope.

%%%%%%%%%%%%%%%%%%%%%%%%%%%%%%%%%%%%%%%%%%%%%%%%%%%%%%%%%%%%%%%%%%%%%%%%%%%%%%%%
\subsection{Numerical methodology} \label{sec:methods_code}
%%%%%%%%%%%%%%%%%%%%%%%%%%%%%%%%%%%%%%%%%%%%%%%%%%%%%%%%%%%%%%%%%%%%%%%%%%%%%%%%

We use our nested grid code \citep{MachidaMatsumoto2012} with $64 \times 64 \times 64$ cells per grid level.
We use the index ``$l$'' to describe a grid level. 
The grid size $L_l$ and cell width $h_l$ of the $l$th grid are twice larger than those of ($l+1$)th grid (e.g., $L_{l+1} = L_l/2$ and $h_{l+1} = h_l/2$).
The base grid ($l = 1$) has the grid size $L_1 = 2\times10^5\,$au and the cell size $h_1 = 3.1\times10^3\,$au, respectively.
Refinement proceeds until the Jeans wavelength is resolved by at least 32 cells.
The finest grid ($l = 14$) has the grid size $L_{14} = 24\,$au and the cell size $h_{14} = 0.38\,$au, respectively.

To allow long-term simulations at reasonable computational cost, we implement a sink cell method with a threshold density $n_{\rm sink}=10^{13}\,\cc$ and a sink radius $r_{\rm sink}=0.5\,$au.
When the gas density exceeds $n_{\rm sink}$, we define that moment as protostar formation ($\tps=0\,$yr) and replace the cell with a sink cell.
After protostar formation, any gas with $n > n_{\rm sink}$ inside $r_{\rm sink}$ is removed and added to the protostellar mass.
The momentum and angular momentum of the gas accreting on the protostar are removed from the calculation domain.

%%%%%%%%%%%%%%%%%%%%%%%%%%%%%%%%%%%%%%%%
\begin{deluxetable}{crrr}
\tablenum{1}
\tablecaption{Model parameters}
\tablewidth{0pt}
\tablehead{
    \colhead{Model} &
    \colhead{$\mu_0$} &
    \colhead{$\theta_0$} &
    \colhead{$\tps$} \\
    \colhead{} &
    \colhead{} &
    \colhead{($^\circ$)} &
    \colhead{($10^4\,$yr)}
}
\decimalcolnumbers
\startdata
T00 & $3$ &  0 & $3$ \\
T10 & $3$ & 10 & $1$ \\
T15 & $3$ & 15 & $5$ \\
T20 & $3$ & 20 & $1$ \\
T30 & $3$ & 30 & $10$ \\
T40 & $3$ & 40 & $1$ \\
T45 & $3$ & 45 & $10$ \\
T50 & $3$ & 50 & $1$ \\
T60 & $3$ & 60 & $10$ \\
T70 & $3$ & 70 & $1$ \\
T75 & $3$ & 75 & $10$ \\
T80 & $3$ & 80 & $1$ \\
T90 & $3$ & 90 & $10$ \\
\hline 
M2 & $2$ & 0 & $1$ \\
M3 & $3$ & 0 & $1$ \\
M4 & $4$ & 0 & $1$ \\
\hline 
B0 & $\infty$ & 0 & $10$
\enddata
\tablecomments{
Column 1: model name;
Column 2: parameter $\mu_0$, the mass-to-flux ratio imposed on the initial gas cloud;
Column 3: parameter $\theta_0$, the angle between the rotation axis and the magnetic field of the initial cloud;
Column 4: end time of the simulation after protostar (sink cell) formation.
}
\label{tab:models}
\end{deluxetable}
%%%%%%%%%%%%%%%%%%%%%%%%%%%%%%%%%%%%%%%%

%%%%%%%%%%%%%%%%%%%%%%%%%%%%%%%%%%%%%%%%%%%%%%%%%%%%%%%%%%%%%%%%%%%%%%%%%%%%%%%%
\subsection{Initial condition and model parameters} \label{sec:methods_models}
%%%%%%%%%%%%%%%%%%%%%%%%%%%%%%%%%%%%%%%%%%%%%%%%%%%%%%%%%%%%%%%%%%%%%%%%%%%%%%%%

We adopt an initial density profile of $f n_{\rm BE}(r)$, where $f=2.1$ is a density enhancement factor to promote cloud contraction and $n_{\rm BE}(r)$ is a critical Bonnor-Ebert density profile \citep{Bonnor1956, Ebert1955}.
The initial central density is $n_0(0) = 5\times10^5\,\cc$, and the total mass and radius of the cloud are $M_0 = 1.23\,\msun$ and $R_0 = 6.21 \times 10^3\,$au, respectively. 
We set a temperature of $T_0 = 10\,$K and a rigid rotation of $\Omega_0 = 1.46 \times 10^{-13}\,$s$^{-1}$.
Under these conditions, the ratios of thermal and rotational energies to the gravitational energy are $\alpha_0 = 0.4$ and $\beta_0 = 0.02$, respectively.

To investigate the impact of magnetic fields on the velocity distribution, we vary two parameters of the initial cloud: (1) the mass-to-flux ratio $\mu_0$\footnote{The mass-to-flux ratio $\mu_0$ is normalized by its critical value, $(4 \pi^2 G)^{-1/2}$.} (or equivalently the initial magnetic field strength $B_0$) and (2) the angle between the rotation axis and the magnetic field direction, $\theta_0$.
Table~\ref{tab:models} lists the models:
\begin{itemize}
    \item Reference model (B0): $\mu_0 = \infty$ ($B_0 = 0\,$G) and $\theta_0 = 0^\circ$.
    \item Models for $\mu_0$ dependence (M2, M3, M4): $\mu_0$ = 2, 3, and 4 ($B_0 = 7.2 \times 10^{-5}, 4.9 \times 10^{-5}$, and $3.5 \times 10^{-5}\,$G) and $\theta_0 = 0^\circ$.
    \item Models for $\theta_0$ dependence (T00 -- T90): $\mu_0 = 3$ ($B_0 = 4.9 \times 10^{-5}\,$G) and $\theta_0 = 0-90^\circ$ in $10^\circ$ increments.
\end{itemize}
$\tps$ in Table~\ref{tab:models} indicates the elapsed time after protostar formation at the end of each simulation.
Due to computational resource limitations, $\tps$ differs among models.
We run simulations up to $\tps = 10^5\,$yr for T30, T45, T60, T75, and T90.
For T00 and T15, we stop at $\tps = 3 \times 10^4\,$yr and $5 \times 10^4\,$yr, respectively, due to the high computational cost.

%%%%%%%%%%%%%%%%%%%%%%%%%%%%%%%%%%%%%%%%%%%%%%%%%%%%%%%%%%%%%%%%%%%%%%%%%%%%%%%%
\subsection{Analysis methodology}\label{sec:methods_analyze}
%%%%%%%%%%%%%%%%%%%%%%%%%%%%%%%%%%%%%%%%%%%%%%%%%%%%%%%%%%%%%%%%%%%%%%%%%%%%%%%%

%%%%%%%%%%%%%%%%%%%%%%%%%%%%%%%%%%%%%%%%
\begin{figure}[t!]
\includegraphics[width=1.0\linewidth]{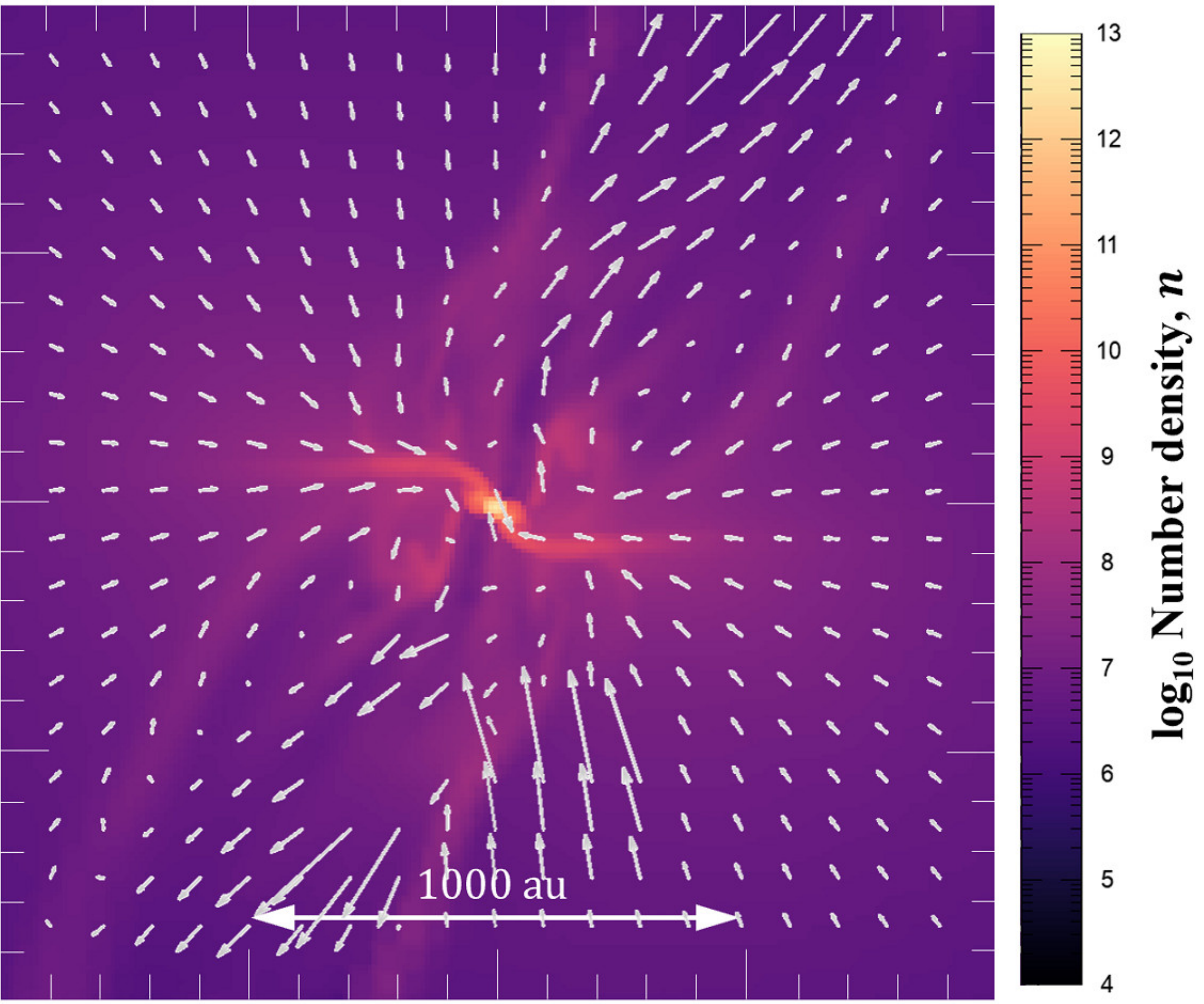}
\caption{
Cross-sectional edge-on view ($y=0$) of gas number density of circumstellar disk for model T30 at $\tps = 5000\,$yr after protostar formation.
The box size is 2000\,au.
The arrows show the gas velocity field.
}
\label{fig:map2d_T30}
\end{figure}
%%%%%%%%%%%%%%%%%%%%%%%%%%%%%%%%%%%%%%%%

We analyze the simulation results to study the velocity structures in the infalling-rotating envelope and the Keplerian disk.
When the misalignment angle, one of our model parameters, is $\theta_0 > 0$, the disk and envelope warp and deform over time as seen in Figure~\ref{fig:map2d_T30} \citep[e.g.,][]{HiranoMachida2019, Hirano2020}.
This implies that the rotation axis of the circumstellar material changes with both distance from the protostar $r$ and with time $\tps$.
Because of this complex rotational structure, we adopt the following simplified procedure:
\begin{enumerate}
    \item We set up radial distance bins centered on the protostar, dividing one order of magnitude in radius into 20 logarithmic intervals $dr$.
    \item For each three-dimensional cell, we decompose its velocity vector into normal (radial) and tangential (rotational) components with respect to the protostar.
    \item For cells whose distances fall within $r$ to $r+dr$, we compute density-weighted averages of the normal and tangential velocities, treating the system as spherically symmetric in one dimension.
    \item We regard the averaged tangential and normal velocities at each radius $r$ as the rotation and infall velocities, $\vrot(r)$ and $\vinfall(r)$, respectively.
\end{enumerate}
Although this procedure includes the outflow regions in the velocity calculation, their contribution is negligible because the density is much lower than in the disk and envelope regions at the same radius bin.
To confirm this, we compared the averaged velocities with the individual cell velocities before averaging, confirming that we are extracting the high-density disk and envelope information.

We calculate the free-fall and Keplerian velocities using (1) the total mass within a given radius, $\Mps + \Menc(r)$, where $\Menc(r)$ is the mass contained within a spherical radius $r$ outside the sink cell, and (2) only the protostellar mass, $\Mps$.
When using the total mass, we denote the velocities with the subscript ``tot'',
\begin{eqnarray}
    \vfftot(r) = \sqrt{\frac{2 G (\Mps + \Menc(r))}{r}}, \label{eq:vfftot} \\
    \vkeplertot(r) = \sqrt{\frac{G (\Mps + \Menc(r))}{r}}, \label{eq:vkeplertot}
\end{eqnarray}
whereas those derived using only the central protostar (or sink) mass, $\Mps$, are denoted with the subscript ``ps'',
\begin{eqnarray}
    \vffps(r) = \sqrt{\frac{2 G \Mps}{r}}, \label{eq:vffps} \\
    \vkeplerps(r) = \sqrt{\frac{G \Mps}{r}}. \label{eq:vkeplerps}
\end{eqnarray}

In Section~\ref{sec:results}, we employ Equations~\ref{eq:vfftot} and \ref{eq:vkeplertot} to evaluate the ratios of the infall and rotation velocities to the free-fall and Keplerian velocities.
In section~\ref{sec:discussion}, we employ Equations~\ref{eq:vffps} and \ref{eq:vkeplerps} to evaluate how $\Menc$ could affect the protostellar mass estimate.

%%%%%%%%%%%%%%%%%%%%%%%%%%%%%%%%%%%%%%%%
\begin{figure*}[p]
    \centering
    \includegraphics[width=1.0\linewidth]{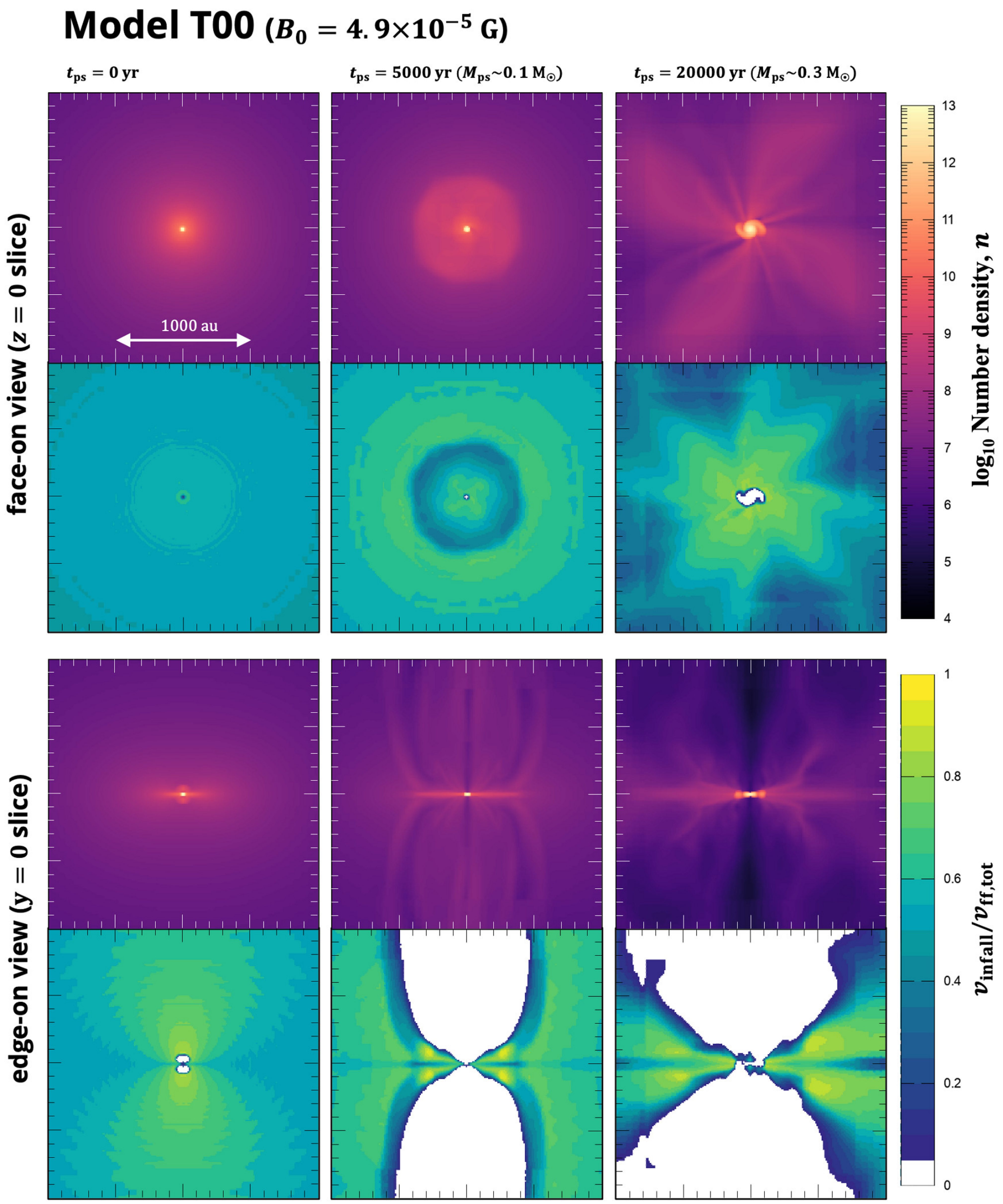}
\caption{
Cross-sectional views of the circumstellar disk at $z=0$ (face-on) and $y=0$ (edge-on) around the protostar for model T00 at $\tps = 0$, 5000, and 20000\,yr after protostar formation: (1) gas number density and (2) the ratio of the infall velocity to the free-fall velocity (Equation~\ref{eq:vfftot}).
The ratio of the infall to the free-fall velocities ($\vinfall/\vfftot > 0$) is color-coded on the color-bar, distinguishing from escape motions ($\vinfall/\vfftot < 0$; white).
The box size is 2000\,au.
}
\label{fig:2dmap_T00}
\end{figure*}
%%%%%%%%%%%%%%%%%%%%%%%%%%%%%%%%%%%%%%%%

%%%%%%%%%%%%%%%%%%%%%%%%%%%%%%%%%%%%%%%%
\begin{figure*}[p]
    \centering
    \includegraphics[width=1.0\linewidth]{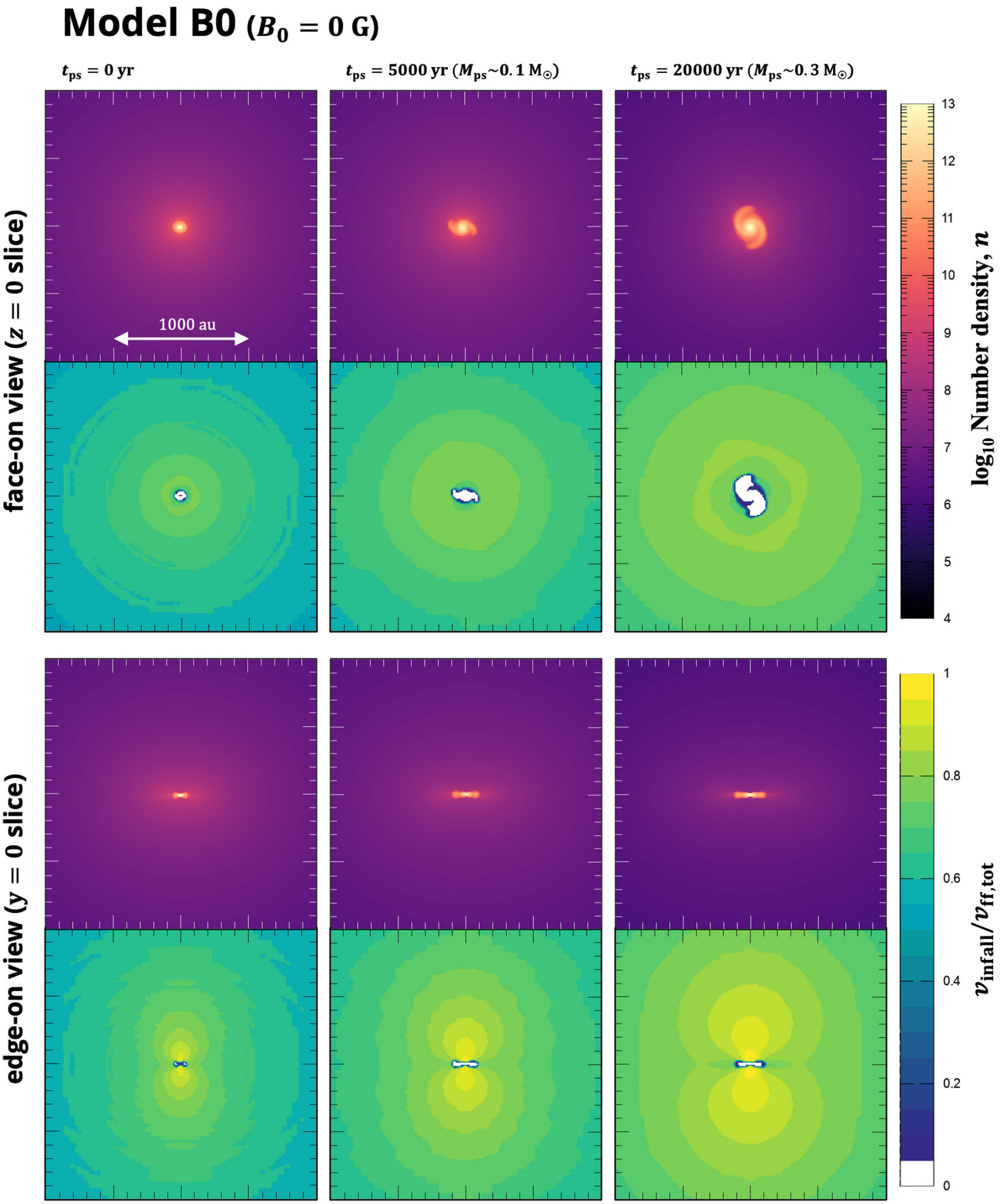}
\caption{
Same as Figure~\ref{fig:2dmap_T00} for the reference model B0.
}
\label{fig:2dmap_B0}
\end{figure*}
%%%%%%%%%%%%%%%%%%%%%%%%%%%%%%%%%%%%%%%%

%%%%%%%%%%%%%%%%%%%%%%%%%%%%%%%%%%%%%%%%
\begin{figure*}[ht!]
    \centering
    \includegraphics[width=0.9\linewidth]{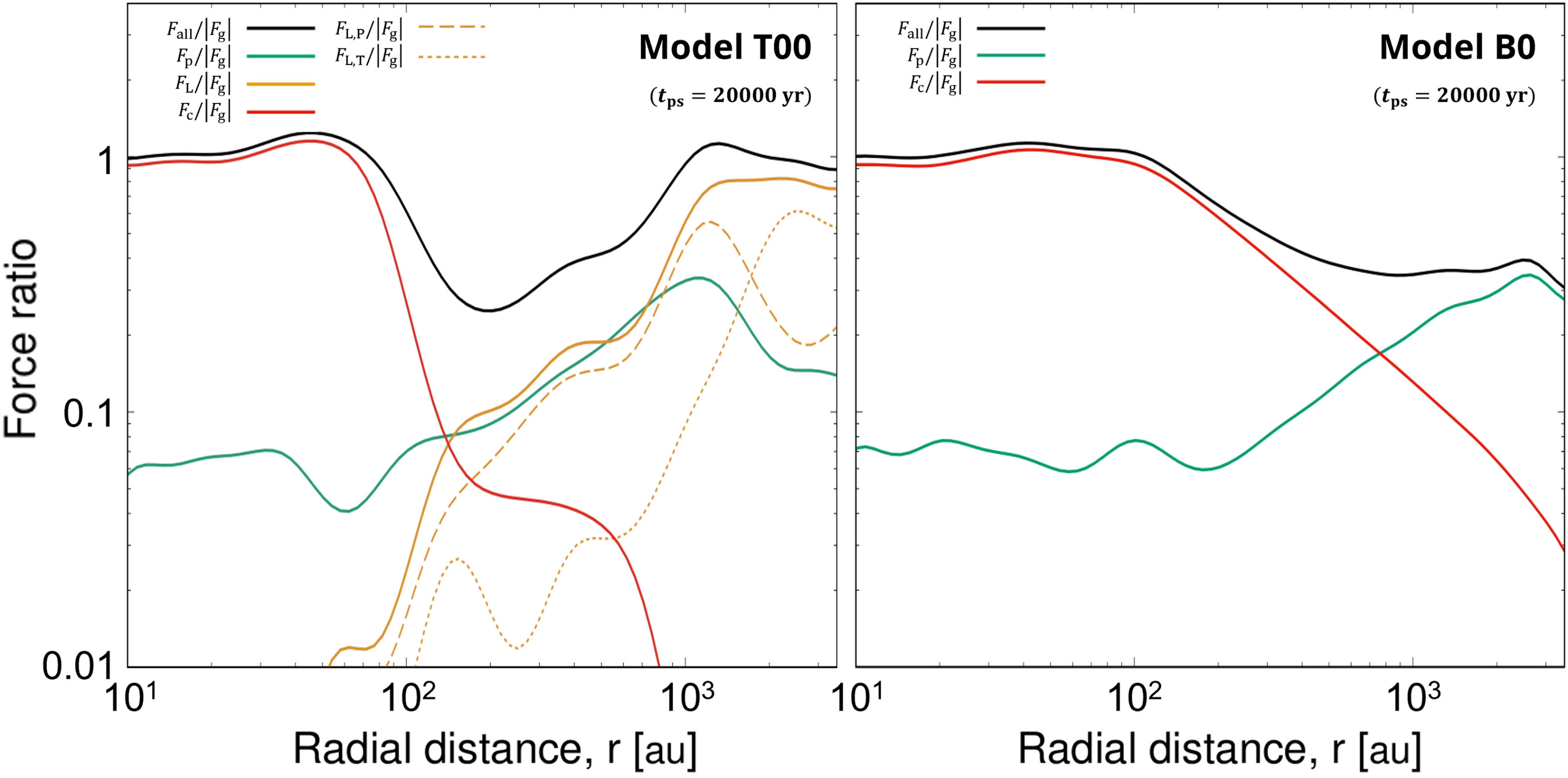}
\caption{
Radial profiles of ratios of all ($F_{\rm all}$), gas pressure gradient ($F_{\rm p}$), Lorentz ($F_{\rm L}$), and centrifugal ($F_{\rm c}$) forces to gravity ($F_{\rm g}$) for models T00 (left) and B0 (right) at $\tps = 2 \times 10^4\,$yr ($\Mps \sim 0.3\,\msun$).
$F_{\rm all}$ is the sum of all except for gravity, $F_{\rm all} = F_{\rm p} + F_{\rm L} + F_{\rm c}$.
The dashed and dotted lines represent ratios of magnetic pressure ($F_{\rm L,P}$) and magnetic tension ($F_{\rm L,T}$) to gravity, respectively, corresponding to the components of the Lorentz force ($F_{\rm L} = F_{\rm L,P} + F_{\rm L,T}$).
Each force is azimuthally averaged.
}
\label{fig:ForceRatio}
\end{figure*}
%%%%%%%%%%%%%%%%%%%%%%%%%%%%%%%%%%%%%%%%

%%%%%%%%%%%%%%%%%%%%%%%%%%%%%%%%%%%%%%%%%%%%%%%%%%%%%%%%%%%%%%%%%%%%%%%%%%%%%%%%
\section{Circumstellar velocity structure} \label{sec:results}
%%%%%%%%%%%%%%%%%%%%%%%%%%%%%%%%%%%%%%%%%%%%%%%%%%%%%%%%%%%%%%%%%%%%%%%%%%%%%%%%

We performed a set of simulations for the models listed in Table~\ref{tab:models} to investigate the velocity structure around Class 0/I protostars.
In this section, we focus on the ratios $\vinfall/\vfftot$ (infall velocity to free-fall velocity) and $\vrot/\vkeplertot$ (rotation velocity to Keplerian velocity) as functions of radius from the protostar.
The free-fall and Keplerian velocities are based on the total mass enclosed within the given radius (Equations~\ref{eq:vfftot} and \ref{eq:vkeplertot}).

%%%%%%%%%%%%%%%%%%%%%%%%%%%%%%%%%%%%%%%%%%%%%%%%%%%%%%%%%%%%%%%%%%%%%%%%%%%%%%%%
\subsection{Distribution of the infall velocity}
%%%%%%%%%%%%%%%%%%%%%%%%%%%%%%%%%%%%%%%%%%%%%%%%%%%%%%%%%%%%%%%%%%%%%%%%%%%%%%%%

Before comparing different models, we first illustrate how magnetic fields influence the infall velocity.
Figures~\ref{fig:2dmap_T00} and \ref{fig:2dmap_B0} show cross-sectional density and velocity ratio ($\vinfall/\vfftot$) distributions for a fiducial magnetized model (T00, $\mu_0=3$) and a non-magnetic reference model (B0, $B_0=0\,$G).
Each figure presents face-on and edge-on views of the disk at $\tps=0$, 5000, and 20000\,yr after protostar (sink cell) formation.

The density distributions (top panels) are consistent with previous studies \citep[e.g.,][]{Hirano2020}.
At $\tps=0\,$yr, the magnetic and non-magnetic models show similar density structures.
In the magnetized model T00, outflows develop by $\tps=5000\,$yr, reducing the density in the polar regions. 
A smaller accretion disk\footnote{During the protostellar accretion phase, gas from the envelope accretes to the accretion disk and thus adds angular momentum to the disk. This angular momentum is transported to the disk’s outer edge, ultimately determining the disk size. We evaluate the disk size by identifying the radius where the radially-averaged velocity abruptly transitions from infall to rotation.} forms near the protostar ($r \lesssim 100\,$au) in T00, while a pseudo-disk supported by magnetic fields appears at outer radii $r \sim 300\,$au.
Over time, both the accretion disk and the pseudo-disk grow in size (the right panels in Figures~\ref{fig:2dmap_T00} and \ref{fig:2dmap_B0}).

Distributions of the velocity ratio (bottom panels in Figures~\ref{fig:2dmap_T00} and \ref{fig:2dmap_B0}) help distinguish infall from outflow.
The velocity ratio $\vinfall/\vfftot$ is shown in color in the regions where the gas is in infall, while the outflow region is in white. 
In the magnetized model T00, the outflow regions where the gas is escaping from the protostar appear white (edge-on view in Figure~\ref{fig:2dmap_T00}).
In contrast, in the non-magnetic model B0, the same region is shown in color, indicating that the gas is infalling.
Within the disk plane (face-on view in Figures~\ref{fig:2dmap_T00} and \ref{fig:2dmap_B0}), the non-magnetic model B0 shows a moderate reduction in infall velocity ($\vinfall/\vfftot \sim 0.8$) due to angular momentum conservation, while the magnetic model T00 experiences an additional slowdown at the pseudo-disk’s outer edge, creating a ring-like region with $\vinfall/\vfftot \sim 0.3$.

When a magnetized cloud collapses, the envelope drags in magnetic field lines, forming an hourglass shape \citep[e.g.,][]{Basu2024}.
If the gas motion is not aligned with the field lines, the Lorentz force decelerates the infall motion.
Figure~\ref{fig:ForceRatio} shows the ratios of various forces (gas pressure gradient $F_{\rm p}$, Lorentz $F_{\rm L}$, and centrifugal force $F_{\rm c}$) to the gravity ($F_{\rm g}$) as functions of radius for models T00 (left panel) and B0 (right panel).
In model T00, the Lorentz force ($F_{\rm L}$) dominates at $r \sim 1000\,$au, where the pseudo-disk forms and the infall velocity stagnates (see Figure~\ref{fig:2dmap_T00}).
This slowly infalling, ring-like region expands over time.

%%%%%%%%%%%%%%%%%%%%%%%%%%%%%%%%%%%%%%%%%%%%%%%%%%%%%%%%%%%%%%%%%%%%%%%%%%%%%%%%
\subsection{Magnetic effect on the infall velocity}
%%%%%%%%%%%%%%%%%%%%%%%%%%%%%%%%%%%%%%%%%%%%%%%%%%%%%%%%%%%%%%%%%%%%%%%%%%%%%%%%

\begin{figure}[t!]
\includegraphics[width=1.0\linewidth]{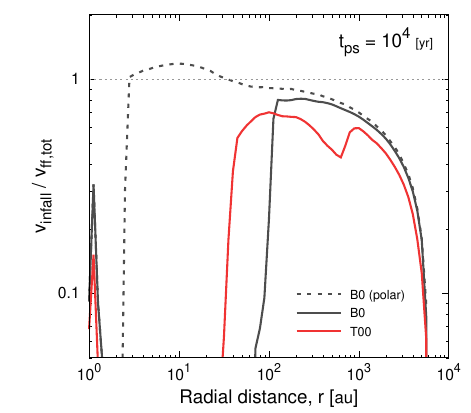}
\caption{
Radial profiles of the velocity ratio, $\vinfall/\vfftot$, around the protostar for models T00 (red line) and B0 (black solid) at $\tps = 10^4\,$yr.
The dashed line represents the velocity ratio of gas inflowing from the polar direction toward the circumstellar disk for model B0, where no outflow appears.
The dotted horizontal line indicates $\vinfall = \vfftot$.
}
\label{fig:VinfallVff_fiducial}
\end{figure}

Figure~\ref{fig:VinfallVff_fiducial} shows the radial profiles of $\vinfall/\vfftot$ at $\tps=10^4\,$yr, using mass-weighted averaging at each radius.
In a classical Shu's solution \citep{Shu1977}, $\vinfall=\vff$ holds within a region where the rarefaction wave passes,
\begin{eqnarray}
    \rrw = \cs \tps \simeq 0.04\,{\rm au} \left( \frac{\tps}{\rm yr} \right) \ , \label{eq:Rrw}
\end{eqnarray}
where $\cs \simeq 0.2\,\kms$ is the sound speed for the gas temperature $10\,$K.
The propagation of the rarefaction wave should be considered when we estimate the free fall velocity \citep[for reference, see Figure~A1 of][]{MachidaNakamura2015}.
At $\tps=10^4\,$yr, $\rrw \sim 400\,$au, and in the non-magnetic model B0, we find $\vinfall/\vfftot \sim 0.8$ within $\rrw$.\footnote{Even without a magnetic field, $\vinfall/\vfftot$ never reaches unity because angular momentum or rotation can not be ignored. Only the polar direction (dashed line in Figure~\ref{fig:VinfallVff_fiducial}) yields $\vinfall/\vfftot \sim 1$.}

In contrast, the magnetized model T00 shows a lower velocity ratio, primarily due to the deceleration at the outer edge of the pseudo-disk (around $r\sim800\,$au, see Figure~\ref{fig:2dmap_T00}).
At the outer edge of the Keplerian disk ($r\sim50\,$au), as the gas flows inwards, the gas transitions from infall-dominated to rotation-dominated motion, causing a sharp drop in $\vinfall/\vfftot$.

These results confirm that magnetic fields suppress infall velocities.
Figure~\ref{fig:ForceRatio} shows the force ratios of magnetic pressure gradient ($F_{\rm L,P}$) and magnetic tension ($F_{\rm L,T}$) forces to gravity.
The magnetic pressure gradient force dominates around the pseudo-disk region, while the magnetic tension force becomes more significant at larger radii.
Consequently, the differences in $\vinfall/\vfftot$ between models T00 and B0 (Figure~\ref{fig:VinfallVff_fiducial}) are interpreted as follows: for $r>1000\,$au, the deceleration in infall velocity is mainly due to the magnetic tension force originating from the initial configuration of the magnetic field lines permeating the gas cloud. In contrast, the sharp drop observed at $r\sim800\,$au is attributable to the magnetic pressure gradient force acting around the outer edge of the pseudo-disk.
The degree of suppression depends on the initial cloud properties and the elapsed time since protostar formation, as described in the following subsections.

%%%%%%%%%%%%%%%%%%%%%%%%%%%%%%%%%%%%%%%%
%%%%%%%%%%%%%%%%%%%%%%%%%%%%%%%%%%%%%%%%

%%%%%%%%%%%%%%%%%%%%%%%%%%%%%%%%%%%%%%%%
\begin{figure}[t!]
\includegraphics[width=1.0\linewidth]{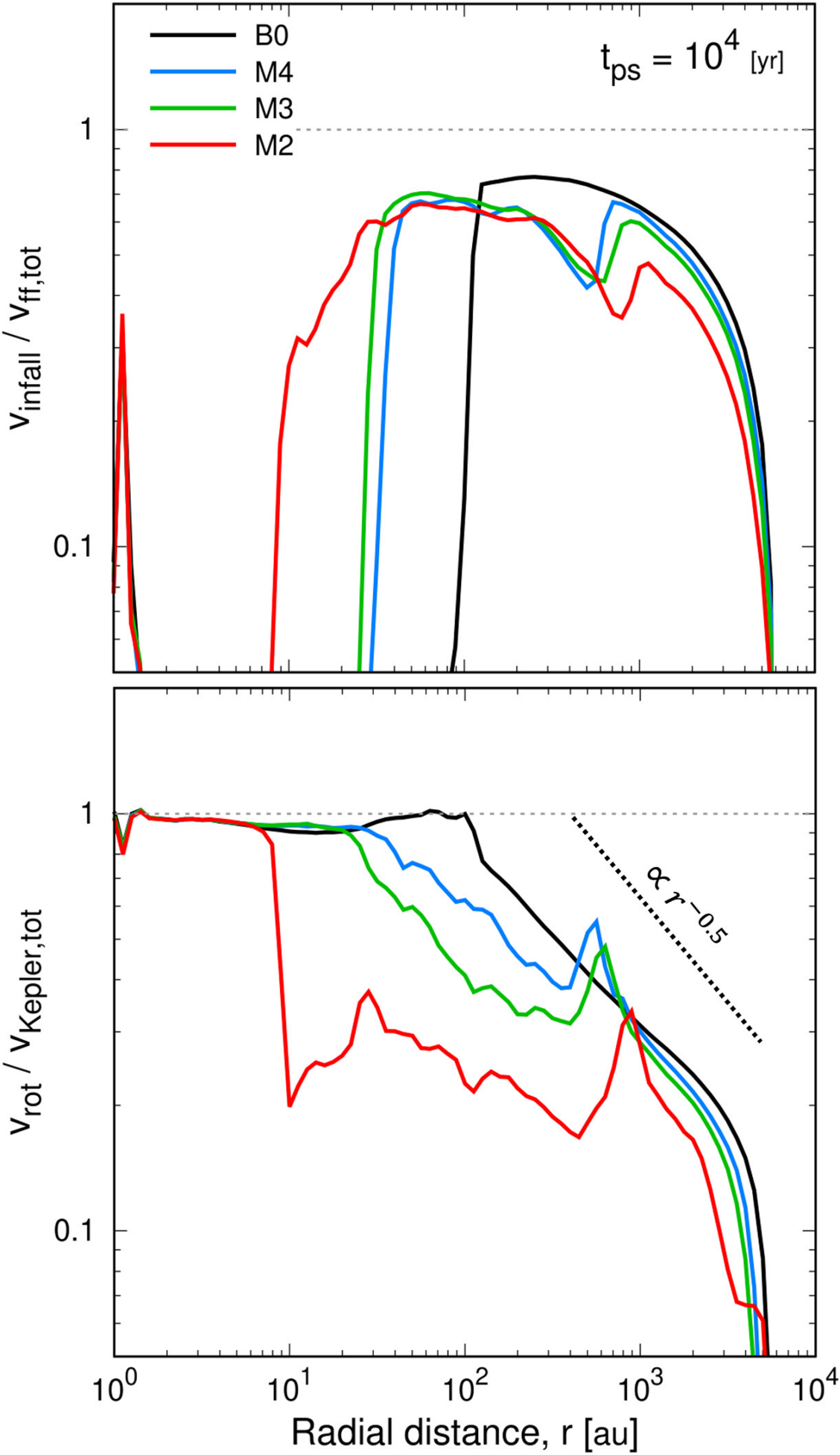}
\caption{
Radial profiles of the velocity ratios, $\vinfall/\vfftot$ (top) and $\vrot/\vkeplertot$ (bottom) at $\tps = 10^4\,$yr for models with different $\mu_0$, M2, M3, M4, and B0 with $\mu_0 = 2$, 3, 4, and $\infty$ ($B_0 = 0\,$G).
The dotted diagonal line in the bottom panel represents a slope of 0.5 with respect to the radius ($\vrot/\vkeplertot \propto r^{-0.5}$), corresponding to the change in the ratio expected when angular momentum is conserved ($\vrot \propto r^{-1}$).
}
\label{fig:rate_mu}
\end{figure}
%%%%%%%%%%%%%%%%%%%%%%%%%%%%%%%%%%%%%%%%

%%%%%%%%%%%%%%%%%%%%%%%%%%%%%%%%%%%%%%%%
\begin{figure}[t!]
\includegraphics[width=1.0\linewidth]{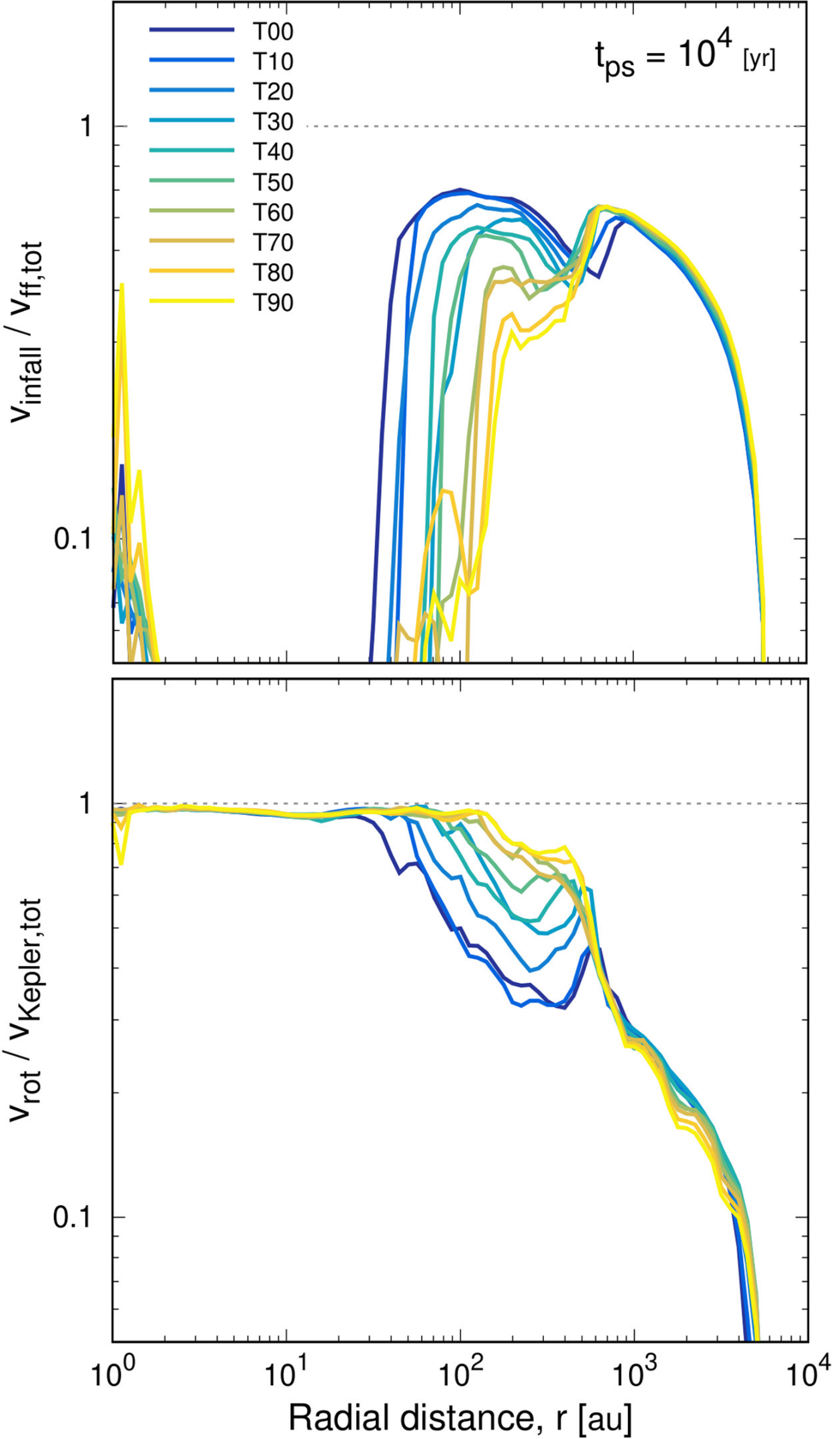}
\caption{
Similar to Figure~\ref{fig:rate_mu} but for models with different $\theta_0$, T00--T90 with $\theta_0 = 0-90^\circ$ in $10^\circ$ increments.
}
\label{fig:rate_theta}
\end{figure}
%%%%%%%%%%%%%%%%%%%%%%%%%%%%%%%%%%%%%%%%

%%%%%%%%%%%%%%%%%%%%%%%%%%%%%%%%%%%%%%%%
\begin{figure}[t!]
\includegraphics[width=1.0\linewidth]{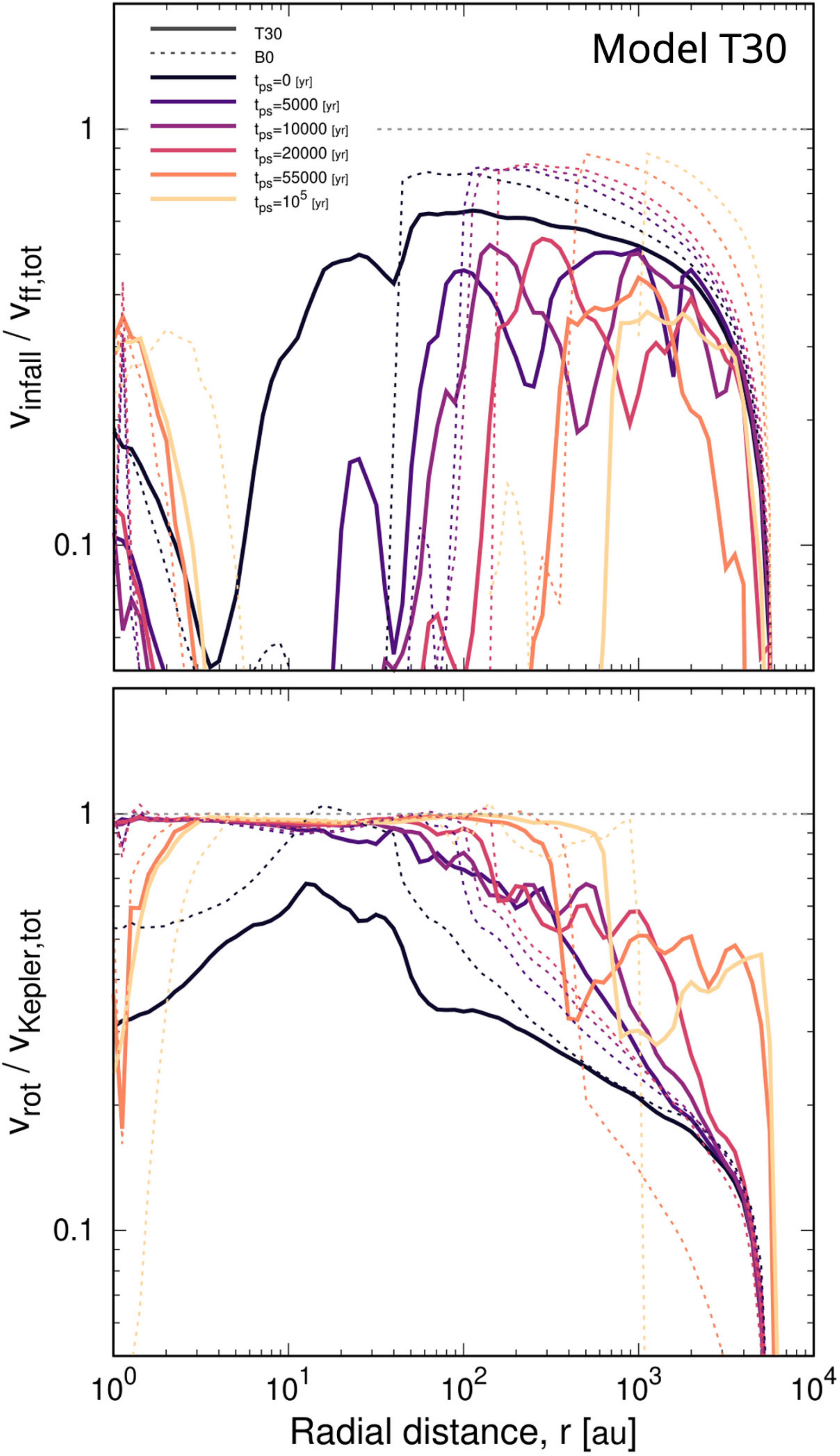}
\caption{
Similar to Figure~\ref{fig:rate_mu} but at epochs of $\tps = 0$, 5000, 10000, 20000, 55000, and $10^5\,$yr for models T30 (solid) and B0 (dotted).
}
\label{fig:rate_time}
\end{figure}
%%%%%%%%%%%%%%%%%%%%%%%%%%%%%%%%%%%%%%%%

%%%%%%%%%%%%%%%%%%%%%%%%%%%%%%%%%%%%%%%%
\begin{figure*}[t!]
\includegraphics[width=1.0\linewidth]{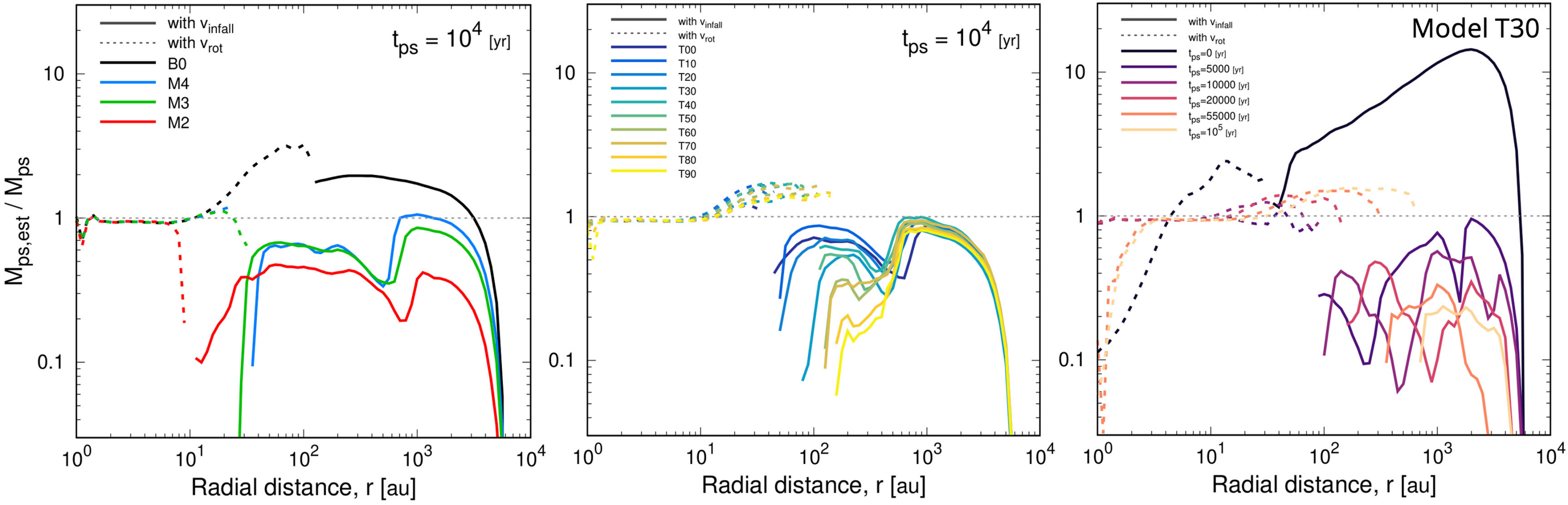}
\caption{
Radial profiles of the ratio of the estimated protostellar mass ($\Mpsest$) using the infall velocity ($\vinfall$; solid line) and the rotation velocity ($\vrot$; dashed) to the true protostellar mass ($\Mps$).
The ratio of the mass estimated using the rotation velocity is plotted for the inner region of the Keplerian disk, while the ratio of the mass estimated using the infall velocity is plotted for the outer region.
The left, center, and right panels correspond to the same models and epochs as in Figures~\ref{fig:rate_mu}, \ref{fig:rate_theta}, and \ref{fig:rate_time}, respectively.
}
\label{fig:Mps_est-true}
\end{figure*}
%%%%%%%%%%%%%%%%%%%%%%%%%%%%%%%%%%%%%%%%

%%%%%%%%%%%%%%%%%%%%%%%%%%%%%%%%%%%%%%%%%%%%%%%%%%%%%%%%%%%%%%%%%%%%%%%%%%%%%%%%
\subsection{Dependence on cloud properties: $\mu_0$ and $\theta_0$} \label{sec:results_param}
%%%%%%%%%%%%%%%%%%%%%%%%%%%%%%%%%%%%%%%%%%%%%%%%%%%%%%%%%%%%%%%%%%%%%%%%%%%%%%%%

Figure~\ref{fig:rate_mu} shows the radial profiles of the velocity ratios at $\tps=10^4\,$yr for models with varying magnetic field strengths.
The key feature is that at the outer edge of the pseudo-disk, $r\sim500-800\,$au, $\vinfall$ decreases while $\vrot$ increases as the radius decreases \citep[e.g.,][]{Machida2005, Machida2006}.
The ratio of the infall velocity to the free-fall velocity ($\vinfall/\vfftot$; top panel) sharply drops at the outer edge of the pseudo-disk, which becomes more significant with higher initial magnetic fields (smaller $\mu_0$).
At the same radius, the ratio of the rotation velocity to the Keplerian velocity ($\vrot/\vkeplertot$; bottom) increases, indicating a shift from infall-dominated to rotation-dominated motion.
Inside the pseudo-disk, $\vinfall/\vfftot$ increases inwards; the gas continues the radial infall until it reaches the Keplerian disk ($\vrot/\vkeplertot \sim 1$).

As the star-forming cloud core contracts, the rotation velocity of the infalling gas scales inversely with radius ($\vrot \propto r^{-1}$) due to the angular momentum conservation.
When normalized by the Keplerian rotation velocity, this relationship becomes $\vrot/\vkeplertot \propto r^{-0.5}$ (dotted diagonal line in Figure~\ref{fig:rate_mu}).
The non-magnetic model B0 follows this correlation.
In the magnetized models M4 and M3, the rotation velocity ratio sharply increases inwards at the outer edge of the pseudo-disk and then decreases over a few 100\,au region where infall velocity rapidly increases.
Further inside, $\vrot/\vkeplertot$ increases inwards until it reaches $\sim 1$, with the radial slope slightly shallower than that expected purely from angular momentum conservation.
On the other hand, in model M2, which has the strongest initial magnetic field among our models, the rotation velocity ratio remains low even at the radius right outside the Keplerian disk ($\vrot/\vkeplertot \sim 0.2-0.3$), indicating efficient angular momentum extraction due to magnetic braking \citep[e.g.,][]{Lam2019, Commercon2022}.
Figure~\ref{fig:rate_mu} shows that a stronger initial magnetic field results in a smaller Keplerian disk radius due to the angular momentum transport by the magnetic fields.

Figure~\ref{fig:rate_theta} shows the velocity ratios in models varying the misalignment angle $\theta_0$ between the rotation axis and the magnetic field.
As in Figure~\ref{fig:rate_mu}, the infall velocity sharply drops inwards at the outer edge of the pseudo-disk.
The increase of the infall velocity over the inner a few 100\,au region is, however, less significant in the models with larger $\theta_0$, 
resulting in generally lower $\vinfall/\vfftot$ values.
Thus, larger misalignment angles yield lower infall velocity in the infalling envelope.

We summarize the effect of the misalignment angle $\theta_0$ on the infall velocity as follows.
Larger misalignment angles ($\theta_0$) result in less effective magnetic breaking \citep[e.g.,][]{Joos2012, Hirano2020}.
Then the Keplerian disk becomes larger for models with larger $\theta_0$, although the radius of the pseudo-disk becomes slightly smaller (see Figure~\ref{fig:rate_theta}, top panel). 
Similarly, in the models with larger $\theta_0$, more angular momentum is retained in the pseudo-disk, leading to faster rotation and a significantly lower infall velocity.
In summary, as $\theta_0$ increases, the spatial extent in which $\vinfall/\vfftot$ decreases becomes smaller, while the rate of reduction of $\vinfall/\vfftot$ increases.

%%%%%%%%%%%%%%%%%%%%%%%%%%%%%%%%%%%%%%%%%%%%%%%%%%%%%%%%%%%%%%%%%%%%%%%%%%%%%%%%
\subsection{Dependence on elapsed time, $\tps$} \label{sec:results_time}
%%%%%%%%%%%%%%%%%%%%%%%%%%%%%%%%%%%%%%%%%%%%%%%%%%%%%%%%%%%%%%%%%%%%%%%%%%%%%%%%

Figure~\ref{fig:rate_time} shows the time evolution of the velocity structure in model T30 over $10^5\,$yr after the protostar formation.
We plot radial profiles at six epochs: $\tps=0$, 5000, 10000, 20000, 55000, and $10^5\,$yr, with corresponding protostellar masses of $\Mps=0.008$, 0.145, 0.249, 0.443, 1.005, and $1.367\,\msun$.
The non-magnetic model B0 is also shown for comparison (dotted lines).

At the very early stage, just after the protostar formation, the maximum value of $\vinfall/\vfftot$ is slightly smaller in model T30 than in model B0.
Nevertheless, both models share the same general trend: outside the radius of $\gtrsim$ a few tens of au, the infall velocity increases inwards.
As time passes in model T30, a pseudo-disk forms and expands outward, and a sharp drop in $\vinfall/\vfftot$ appears at its outer edge.
In contrast, no such decrease in $\vinfall/\vfftot$ is observed in the envelope of model B0.
By the end of the simulation at $\tps = 10^5\,$yr, the infall velocity ratio in the envelope is about $\vinfall/\vfftot \sim 0.5$ in model T30.
Consequently, the mass estimated using the infall velocity alone is only $0.5^{2} = 0.25$ times the actual mass.

Comparison with observational timescales further supports the applicability of our results.
Among the sources discussed in Section~\ref{sec:intro}, L1489~IRS with an evolutionary timescale of $(3-8)\times10^4\,$yr \citep{Sai2022} shows a $\vinfall/\vfftot\sim0.4$ whereas L1527~IRS with $<3\times10^5\,$yr \citep{Tobin2012} shows $\vinfall/\vfftot\sim0.3$.
Figure~\ref{fig:rate_time} shows an infall-to-free-fall ratio of $\vinfall/\vfftot\sim0.3-0.4$ at $\tps=55000$ and $10^5\,$yr, which are consistent with observational estimates.

%%%%%%%%%%%%%%%%%%%%%%%%%%%%%%%%%%%%%%%%%%%%%%%%%%%%%%%%%%%%%%%%%%%%%%%%%%%%%%%%
\section{Estimation of the protostellar mass} \label{sec:discussion}
%%%%%%%%%%%%%%%%%%%%%%%%%%%%%%%%%%%%%%%%%%%%%%%%%%%%%%%%%%%%%%%%%%%%%%%%%%%%%%%%

We investigate uncertainties in the mass estimation using the protostellar mass ($\Mps$) and the circumstellar rotation and infall velocities ($\vrot$ and $\vinfall$) obtained from our simulations.
To estimate the protostellar mass, one typically uses:
\begin{eqnarray}
    \Mpsest = && \frac{r\vrot^2}{G} \hspace{7mm} \text{(if $r < \rkepler$)} \ , \label{eq:Mpsest_rot} \\
    && \frac{r\vinfall^2}{2G} \hspace{4.5mm} \text{(if $r > \rkepler$)} \ , \label{eq:Mpsest_infall}
\end{eqnarray}
where $\rkepler$ is the Keplerian disk radius.
Figure~\ref{fig:Mps_est-true} shows the ratio $\Mpsest/\Mps$ for the same models and epochs presented in Figures~\ref{fig:rate_mu}-\ref{fig:rate_time}.
To distinguish between accretion-dominated and rotation-dominated regions, we adopt $\rkepler$ where $\vrot/\vkepler = 0.8$ on the radial profile and calculate $\Mps$ inside $\rkepler$ using Equation~\ref{eq:Mpsest_rot} and outside $\rkepler$ using Equation~\ref{eq:Mpsest_infall}.
Inside the Keplerian disk ($r \lesssim 10\,$au), the estimated mass closely matches the true mass ($\Mpsest/\Mps \simeq 1$).
At the outer edge of the Keplerian disk, the mass estimated from the rotational velocity becomes larger than the actual protostellar mass ($\Mpsest/\Mps > 1$; see below).
In the envelope region, the ratio of mass estimated from the infall velocity to the protostellar mass falls below unity ($\Mpsest/\Mps < 1$), reaching a minimum at the outer edge of the pseudo-disk. 
Beyond the pseudo-disk, the coexistence of two opposing effects -- the previously mentioned velocity ratio ($\fff$), which tends to decrease $\Mpsest/\Mps$, and the enclosed mass ($\Menc$) described below, which tends to increase $\Mpsest/\Mps$ -- results in a ratio of $\Mpsest/\Mps$ approaching unity.
To what extent the two effects cancel out depends on the initial physical parameters of the gas cloud, such as the initial magnetic field strength.
For example, in the strong magnetic field model M2, the enhanced magnetic suppression of the infall velocity results in $\Mpsest/\Mps$ remaining below unity even beyond the pseudo-disk.

By considering the enclosed mass ($\Menc$), we can refine the protostellar mass estimate:
\begin{eqnarray}
    \Mps = && \frac{r(\fkepler \vrot)^2}{G} - \Menc(r) \hspace{4.5mm} \text{(if $r < \rkepler$)}, \label{eq:Mps_rot}\\
    && \frac{r(\fff \vinfall)^2}{2G} - \Menc(r) \hspace{2.5mm} \text{(if $r > \rkepler$)} \ , \label{eq:Mps_infall}
\end{eqnarray}
where velocity ratios are defined as $\fkepler = \vkeplertot/\vrot (\ge 1)$ and $\fff = \vfftot/\vinfall (\ge 1)$.
Comparing Equations~\ref{eq:Mpsest_rot} and \ref{eq:Mpsest_infall} with Equations~\ref{eq:Mps_rot} and \ref{eq:Mps_infall}, we identify two factors of error in the protostellar mass estimation using rotation or infall velocities: (1) deviations in the velocity ratios ($\fkepler$ and $\fff$) from unity due to magnetic fields and rotation, and (2) the contribution of $\Menc$, the mass enclosed within radius $r$ but not part of the protostar.

%%%%%%%%%%%%%%%%%%%%%%%%%%%%%%%%%%%%%%%%
\begin{figure}[t!]
\includegraphics[width=1.0\linewidth]{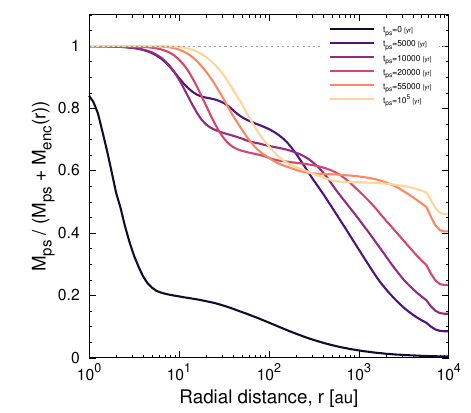}
\caption{
Radial profiles of the ratio of the protostellar mass ($\Mps$) to the total mass ($\Mps + M_{\rm enc}(r)$) for model T30 at $\tps = 0$, 5000, 10000, 20000, 55000, and $10^5\,$yr ($\Mps = 0.008$, 0.145, 0.249, 0.443, 1.005, and $1.367\,\msun$).
}
\label{fig:MpsMenc}
\end{figure}
%%%%%%%%%%%%%%%%%%%%%%%%%%%%%%%%%%%%%%%%

%%%%%%%%%%%%%%%%%%%%%%%%%%%%%%%%%%%%%%%%
\begin{figure}[t!]
\includegraphics[width=1.0\linewidth]{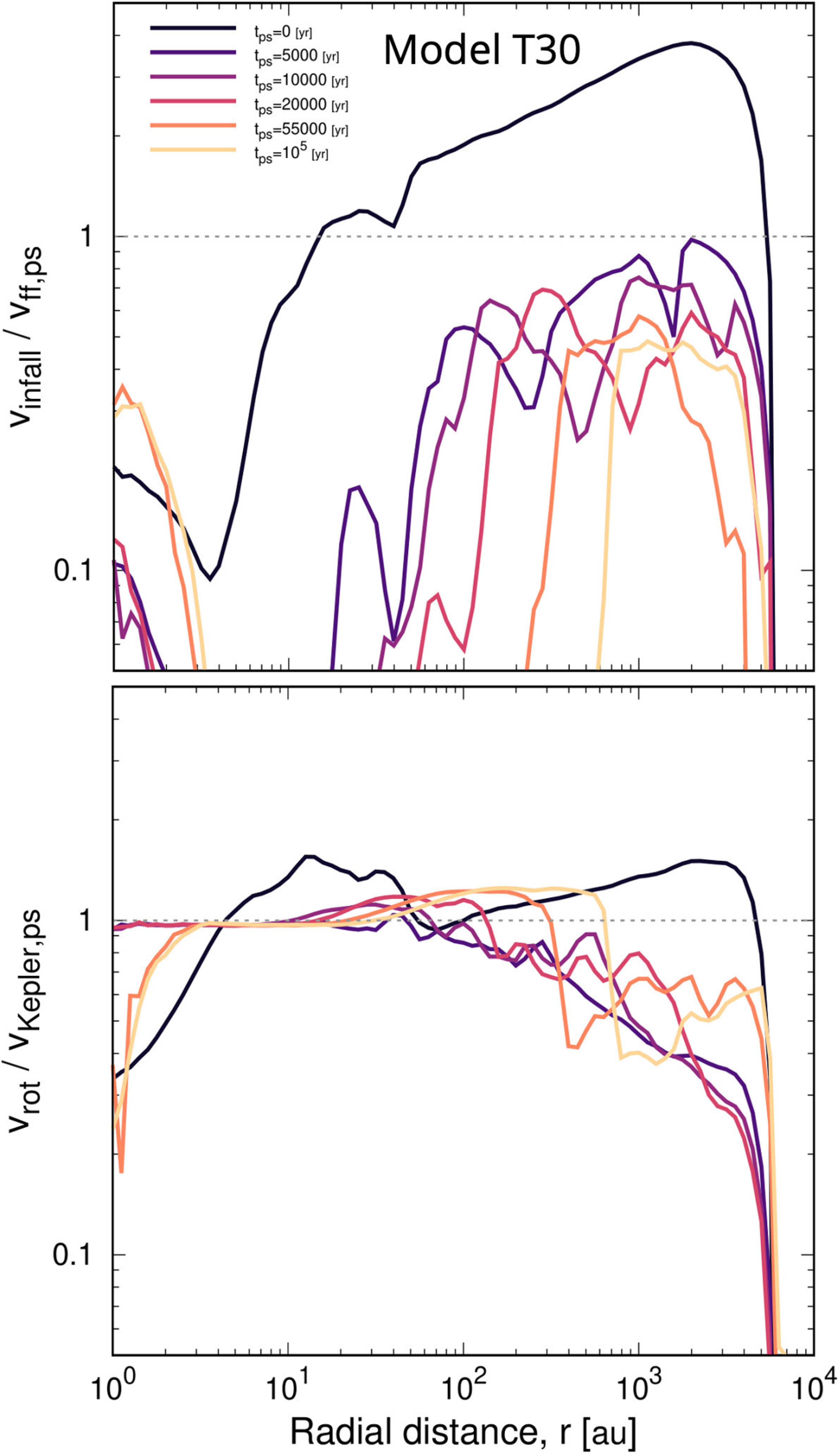}
\caption{
Radial profiles of the velocity ratios, $\vinfall/\vffps$ (top) and $\vrot/\vkeplerps$ (bottom) at $\tps = 0$, 5000, 10000, 20000, 55000, and $10^5\,$yr for model T30.
}
\label{fig:rate_time_MpsOnly}
\end{figure}
%%%%%%%%%%%%%%%%%%%%%%%%%%%%%%%%%%%%%%%%

Figure~\ref{fig:MpsMenc} shows the ratio of the protostellar mass to the total mass, $\Mps / (\Mps + \Menc)$.
Inside the Keplerian disk, the circumstellar mass is negligible compared to the protostellar mass, so $\Mps / (\Mps + \Menc) \sim 1$.
However, near the outer edge of the Keplerian disk, $\Menc$ becomes significant, causing an overestimation of the protostellar mass (Figure~\ref{fig:Mps_est-true}).
At the outer edge of the pseudo-disk, the protostellar mass accounts for only about 60\% of the total mass inside the radius, $\Mps / (\Mps + \Menc) \sim 0.6$.

Significance of $\Menc$ compared with $\Mps$ also depends on the evolutionary stage.
At an early phase, when the protostellar mass is still small ($\Mps < \Menc$; see Figure~\ref{fig:MpsMenc}), the large fraction of circumstellar mass tends to increase the estimated protostellar mass (right panel of Figure~\ref{fig:Mps_est-true}).
This bias arises directly from the second term in Equations~\ref{eq:Mps_rot} and \ref{eq:Mps_infall}, where neglecting $\Menc$ leads to overestimation of the protostellar mass.

Equations~\ref{eq:Mps_rot} and \ref{eq:Mps_infall} illustrate a delicate balance; ignoring the velocity ratios ($\fkepler$ and $\fff (\geq1)$) underestimates $\Mps$, while neglecting the circumstellar mass ($M_{\rm enc}$) overestimates $\Mps$.
By comparing $\vinfall/\vffps$ and $\vrot/\vkeplerps$ (Figure~\ref{fig:rate_time_MpsOnly}) with $\vinfall/\vfftot$ and $\vrot/\vkeplertot$ (Figure~\ref{fig:rate_time}), we can clearly see the contribution of $\Menc$.
Inside the Keplerian disk where $\Mps \gg \Menc$, the rotational velocity ratio remains constant, $\vrot/\vkeplerps \sim \vrot/\vkeplertot \sim 1$.
However, outside the Keplerian disk where $\Mps \sim \Menc$, neglecting the influence of the circumstellar mass $\Menc$ brings the infall velocity ratio $\vinfall/\vffps$ closer to unity ($\sim 0.5-0.9$), particularly outside the pseudo-disk, unlike $\vinfall/\vfftot$ (Figure~\ref{fig:rate_time}), which is always much less than unity.
As a result, when we estimate the protostellar mass using the infall velocity of the envelope outside the pseudo-disk, the effect of the magnetic field, which decreases the infall velocity, and the effect of $\Menc$, which increases the infall velocity, cancel with each other to some extent (Figure~\ref{fig:Mps_est-true}).

We identify five distinct regions in the circumstellar environment, illustrated in the bottom panel of Figure~\ref{fig:overview}.
The accuracies of the protostellar mass estimation for each region differ for distinct reasons:
\begin{enumerate}
    \item Outside the pseudo-disk, due to the coexistence of two opposing effects -- the velocity ratio ($\fff$), which tends to decrease $\Mpsest/\Mps$, and the enclosed mass ($\Menc$), which tends to increase $\Mpsest/\Mps$ -- the resulting ratio $\Mpsest/\Mps$ approaches unity.
    \item At the outer edge of the pseudo-disk, the magnetic field (Lorentz force) reduces the gas infall velocity, resulting in high $\fff$ and $\Mpsest/\Mps \sim 0.3$. 
    \item Inside the pseudo-disk, the infall velocity approaches the free-fall velocity and $\fff$ decreases as the radius decreases. The estimated protostellar mass becomes closer to the actual value up to $\Mpsest/\Mps \sim 0.6$.
    \item At the outer edge of the Keplerian disk $\Menc$ is comparable to $\Mps$, which causes an overestimation of the protostellar mass ($\Mpsest/\Mps \sim 2$) using $v_{\rm rot}$, although $\fkepler$ is $\sim 1$.
    \item Inside the Keplerian disk where $\Mps \gg \Menc$, the protostellar mass estimated from $v_{\rm rot}$ agrees reasonably with the protostellar mass, i.e. $\Mpsest/\Mps \simeq 1$.
\end{enumerate}

In summary, the comparison of the actual protostellar mass with the mass estimated from the observed gas velocities is inherently complex.
The difference is time-dependent and is due to the deviation of the infall velocity from the free fall velocity, as well as the enclosed mass (i.e., disk and envelope).
Our results provide a theoretical framework for interpreting observed velocity structures and their implications for mass determinations in Class 0/I protostars.

%%%%%%%%%%%%%%%%%%%%%%%%%%%%%%%%%%%%%%%%%%%%%%%%%%%%%%%%%%%%%%%%%%%%%%%%%%%%%%%%
\section{Conclusion} \label{sec:conclusion}
%%%%%%%%%%%%%%%%%%%%%%%%%%%%%%%%%%%%%%%%%%%%%%%%%%%%%%%%%%%%%%%%%%%%%%%%%%%%%%%%

We performed a set of three-dimensional magnetohydrodynamic simulations to investigate the velocity structure of circumstellar disks and envelopes of Class 0/I protostars.
Our results confirm that the infall velocity of the infalling-rotating envelope is systematically smaller than the free-fall velocity (Figures~\ref{fig:rate_mu}-\ref{fig:rate_time}).
We then compared protostellar masses obtained in our simulations with those estimated using the commonly employed observational method -- based on infall and rotation velocities -- to evaluate the accuracy of mass estimation (Figure~\ref{fig:Mps_est-true}).

We found that the mass estimation accuracy varies by a factor of about $0.3-2$, depending on the distance from the central star where the gas velocities are measured (Figure~\ref{fig:overview}).
While masses estimated from disk rotation velocities near the protostar closely match the true protostellar mass, masses estimated from the infall velocity of the envelope can underestimate the true protostellar mass.
This naturally reproduces the ``factor-of-three problem,'' in which the observed infall velocity of the envelope in Class 0/I protostars is several times smaller than the expected free-fall velocity \citep[see also][]{Tu2024}.

Moreover, we confirmed that $\fff$ (in Equation~\ref{eq:Mps_infall}) depends on the physical conditions of the gas cloud and the evolutionary stage of the protostar.
Specifically, the discrepancy increases with:
(1) a stronger initial magnetic field ($\mu_0$),
(2) a larger angle between the rotation axis and the magnetic field direction ($\theta_0$), and
(3) longer elapsed time ($\tps$) after protostar formation.
If future statistical observations can correlate the global magnetic field strength and orientation in gas clouds with the ratio of protostellar masses derived from infall and rotation velocities, the trends identified in this study can be observationally verified.

%%%%%%%%%%%%%%%%%%%%%%%%%%%%%%%%%%%%%%%%%%%%%%%%%%%%%%%%%%%%%%%%%%%%%%%%%%%%%%%%

\begin{acknowledgments}
We have benefited greatly from discussions with Yusuke Aso.
We thank our anonymous referee for constructive comments on this study.
This work used the computational resources of the HPCI system provided by the supercomputer system SX-Aurora TSUBASA at Tohoku University Cyber Sciencecenter and Osaka University Cybermedia Center through the HPCI System Research Project (Project ID: hp220003, hp230035, and hp240010) and Earth Simulator at JAMSTEC provided by 2022, 2023, and 2024 Koubo Kadai.
Numerical analyses were carried out on analysis servers at Center for Computational Astrophysics, National Astronomical Observatory of Japan.
This work was supported by JSPS KAKENHI Grant Numbers JP21H01123 and JP21K13960 (S.H.) and JP18H05222, JP20H05847, and 24K00674 (Y.A.).
This work was also supported by a NAOJ ALMA Scientific Research grant (No. 2022-22B). 
\end{acknowledgments}

%%%%%%%%%%%%%%%%%%%%%%%%%%%%%%%%%%%%%%%%%%%%%%%%%%%%%%%%%%%%%%%%%%%%%%%%%%%%%%%%

%\vspace{5mm}
%\facilities{HST(STIS), Swift(XRT and UVOT), AAVSO, CTIO:1.3m,CTIO:1.5m,CXO}

%%%%%%%%%%%%%%%%%%%%%%%%%%%%%%%%%%%%%%%%%%%%%%%%%%%%%%%%%%%%%%%%%%%%%%%%%%%%%%%%

%\software{astropy \citep{2013A&A...558A..33A,2018AJ....156..123A},  
%          Cloudy \citep{2013RMxAA..49..137F}, 
%          Source Extractor \citep{1996A&AS..117..393B}
%          }

%%%%%%%%%%%%%%%%%%%%%%%%%%%%%%%%%%%%%%%%%%%%%%%%%%%%%%%%%%%%%%%%%%%%%%%%%%%%%%%%

%%%%%%%%%%%%%%%%%%%%%%%%%%%%%%%%%%%%%%%%%%%%%%%%%%%%%%%%%%%%%%%%%%%%%%%%%%%%%%%%

%\appendix
%\restartappendixnumbering

%%%%%%%%%%%%%%%%%%%%%%%%%%%%%%%%%%%%%%%%%%%%%%%%%%%%%%%%%%%%%%%%%%%%%%%%%%%%%%%%

\bibliography{ms}{}
\bibliographystyle{aasjournal}

%%%%%%%%%%%%%%%%%%%%%%%%%%%%%%%%%%%%%%%%%%%%%%%%%%%%%%%%%%%%%%%%%%%%%%%%%%%%%%%%



%%%%%%%%%%%%%%%%%%%%%%%%%%%%%%%%%%%%%%%%%%%%%%%%%%%%%%%%%%%%%%%%%%%%%%%%%%%%%%%%

\end{CJK}%! To show the Japanese language.

\end{document}